\documentclass[preprint,aps,pra,showpacs,nofootinbib]{revtex4}
\usepackage{graphicx}
\usepackage{amsmath}
\usepackage{amsfonts}
\usepackage{bm}
\usepackage{color}

\begin{document}

\title[Coulomb glory effect: relativistic theory]{Coulomb glory effect in
collisions of antiprotons with heavy nuclei: relativistic theory}
\date{\today}

\author{A~V~Maiorova$^1$, D~A~Telnov$^1$, V~M~Shabaev$^1$,
V~A~Zaitsev$^1$, G~Plunien$^2$ and T~St\"ohlker$^{3,4}$}
\address{$^1$ Department of Physics, St.~Petersburg State
University, Ulianovskaya 1, Petrodvorets, St.~Petersburg 198504,
Russia}
\address{$^2$ Institut f\"{u}r Theoretische Physik, TU Dresden,
Mommsenstrasse 13, D-01062 Dresden, Germany}
\address{$^3$ Gesellschaft f\"{u}r Schwerionenforschung,
Planckstrasse 1, D-64291 Darmstadt, Germany}
\address{$^4$ Physikalisches Institut, Philosophenweg 12,
 D-69120 Heidelberg, Germany}
%\ead{maiorova@pcqnt1.phys.spbu.ru}

\begin{abstract}
Collisions of antiprotons with bare uranium nuclei are studied
for scattering angles nearby 180$^{\circ}$ in the framework of relativistic
theory. The Coulomb glory phenomenon is investigated at energies of the
antiprotons in the range 100~eV to 2.5~keV. The vacuum polarization effect
and the anomalous magnetic moment of the antiproton are taken into account.
The estimations of possible influence of such effects as radiative
recombination and antiproton annihilation are given.
\end{abstract}
 %\submitto{\jpb}
\maketitle

\section{Introduction}
New facilities for antiproton and ion research at GSI will give an
opportunity to observe the Coulomb glory phenomenon, which was predicted by
Demkov and coauthors \cite{gl01,gl02}. The effect consists in a
prominent maximum of the differential cross section (DCS) in the backward
direction at a certain energy of the incident particle, provided the
interaction with the target is represented by a screened Coulomb attraction
potential.
In our previous papers \cite{Maiorova07,Maiorova08} we investigated the
backward scattering of antiprotons by highly charged and neutral uranium
($Z=92$) in the framework of the non-relativistic quantum mechanics. It was
shown that the Coulomb glory effect takes place due to the screening of the
nuclear Coulomb attraction by the electrons. At the energy where the
strongest effect is predicted, the DCS in the backward direction may exceed
the corresponding Rutherford DCS by several times or even several hundred
times, depending on the number of electrons in the ion \cite{Maiorova07}.
In collisions of antiprotons with the bare uranium, the Coulomb glory is
also present because of the screening properties of the vacuum polarization
(VP) potential. In this case, the effect predicted was not such as large
but still noticeable (about 5\% \cite{Maiorova08}).

In the present paper, we investigate the Coulomb glory effect in collisions
of antiptotons with bare uranium nuclei in the framework of the
relativistic quantum theory. Since the kinetic energy of antiprotons in
the Coulomb glory region is rather low (does not exceed a few keV), one can
expect that the relativistic corrections may not alter the results
significantly. It is certainly the case for the uranium ions where the
effect itself is very large. However, for the bare nuclei the situation is
not that straightforward. Although the ratio of the velocity of antiprotons
to the velocity of light $v_{\bar{p}}/c$ is quite small in the energy
range under consideration ($\le 10^{-3}$), relativistic effects appear
important for the scattering in the backward direction. As was shown in the
paper \cite{Milstein04}, in the low-energy limit the differential cross
section (DCS) of the relativistic Coulomb scattering
$d\sigma^{\mathrm{C}}/d\Omega$ at the angle $\theta=180^{\circ}$ can be
expressed as follows:
\begin{eqnarray}
\frac{d\sigma^{\mathrm{C}}}{d\Omega}\left(\theta=180^{\circ}\right)
=\left[1+\sqrt{2\pi^{3}
\frac{v_{\bar{p}}}{c}\left(\frac{Z}{c}\right)^{3}}\right]
\frac{d\sigma^{\mathrm{B}}}{d\Omega}\left(\theta=180^{\circ}\right)
\label{asimp}
\end{eqnarray}
where $d\sigma^{\mathrm{B}}/d\Omega$ is the DCS obtained in the
(relativistic) first Born approximation. We note that for small
values of the parameter $v_{\bar{p}}/c$, the Born DCS
$d\sigma^{\mathrm{B}}/d\Omega$ is nearly equal to the
non-relativistic Rutherford DCS. According to Eq. (\ref{asimp}),
the contribution to the DCS due to relativistic effects can be as
large as 10\% -- 20\%, depending on the energy of antiprotons in
the interval 100~eV -- 2.5~keV. Of course, any estimation based on
Eq. (\ref{asimp}) is valid for the pure Coulomb potential (that
is, point nucleus charge distribution) only. For the finite size
nucleus, the situation is somewhat different. This especially
concerns the antiproton scattering, where the nuclear size effect
can be rather important. In this case, as our calculations for an
extended nucleus show, the relativistic corrections to the DCS at
the angle $\theta=180^{\circ}$ do not exceed 10\%. Still, they are
quite significant and must be taken into account. In our present
study, the Dirac theory is used to describe collisions of
antiprotons with bare uranium nuclei. Besides the finite nucleus
potential, the equation of motion includes the exact one-loop VP
potential and the interaction term due to the anomalous magnetic
moment of the antiproton. Atomic units ($\hbar=e=m_e=1$) are used
in the paper.

\section{Basic formulas}
In the relativistic quantum theory, the differential cross section of the
scattering of unpolarized beam in the central field is given by the
following equation \cite{Landau4}:
\begin{equation}
\frac{d\sigma}{d\Omega}=|A|^{2}+|B|^{2} \label{dcs}
\end{equation}
where
\begin{eqnarray}
A(\theta)= \frac{1}{2\mathrm{i}
p}\sum_{l=0}^\infty\{(l+1)[\exp(2\mathrm{i}\delta_{l+1/2,\,l})-1]
\nonumber \\
+l[\exp(2\mathrm{i}\delta_{l-1/2,\,l})-1]\}P_{l}(\cos\theta),\label{rjadA}
\end{eqnarray}
\begin{equation}
B(\theta) =
\frac{1}{2p}\sum_{l=1}^\infty[\exp(2\mathrm{i}\delta_{l+1/2,\,
l})-\exp(2\mathrm{i}\delta_{l-1/2,\,l})]P^{1}_{l}(\cos\theta).\label{rjadB}
\end{equation}
Here $p$ is the momentum of the antiproton, $P_{l}(\cos\theta)$ are the
Legendre polynomials and $P^{1}_{l}(\cos\theta)$ are the associated
Legendre functions. The phase shifts $\delta_{j,\,l}$ corresponding to the
total angular momentum $j$ and orbital momentum $l$ can be expressed as a
sum of the phase shifts $\delta_{j,\,l}^{s}$, produced by the short-range
part of the scattering potential, and the Coulomb phase shift
$\delta_{j,\,l}^{c}$:
\begin{equation}
\delta_{j,\,l} = \delta_{j,\,l}^{s}+\delta_{j,\,l}^{c}.\,
\label{phase}
\end{equation}
The Coulomb phase shifts can be represented as follows \cite{Landau4}:
\begin{eqnarray}
\delta_{j,\,l}^{c} = \xi -
\arg\Gamma(\gamma+1-i\nu)-\frac{\pi\gamma}{2}+\frac{\pi l}{2},\,
\label{phase2}
\end{eqnarray}
where $\nu = -Z\varepsilon/cp$ is the Coulomb parameter,
$\varepsilon$ is the total energy of the antiproton,
$\exp(-2\mathrm{i}\xi)=(\gamma+\mathrm{i}\nu)/(\kappa+\mathrm{i}\nu
m_{\bar{p}}c^{2}/\varepsilon)$, $\gamma =
\sqrt{\kappa^{2}-(Z/c)^{2}}$, and the quantum number
$\kappa=(-1)^{j+l+1/2}(j+1/2)$.

We consider antiprotons moving in the central field $V(r)$ and take into
account the anomalous magnetic moment of the antiproton. In this case, the
Dirac equation appears in the following form (see, e.g., \cite{Borie83,pach04}):
\begin{eqnarray}
\left[c(\bm{\alpha}\cdot \bm{p}) + m_{\bar{p}}c^{2}\beta + V(r) -
\frac{\mathrm{i}\kappa_{0}}{2m_{\bar{p}}c}V'(r)
(\bm{\gamma}\cdot\hat{\bm{r}}) \right]\psi(r) =
\varepsilon\psi(r)\,, \label{direq}
\end{eqnarray}
where $\kappa_{0} = 1.792 847 34$ is the anomalous magnetic
moment, $V'(r)$ stands for the first derivative of the potential
$V(r)$, and $\hat{\bm{r}}$ denotes the unit vector in the $\bm{r}$
direction. Eq. (\ref{direq}) has solutions corresponding to the
definite total angular momentum $j$ and its projection $m$:
\begin{eqnarray}
\psi(r) = \frac{1}{r}\left(%
\begin{array}{l}
       G(r)\Omega_{jlm}(\hat{\bm{r}}) \\
    iF(r)\Omega_{jl'm}(\hat{\bm{r}}) \\
\end{array}%
\right).
\end{eqnarray}
Here $\Omega_{jlm}(\hat{\bm{r}})$ are the spherical spinors,
$l=j\pm\frac{1}{2}$, and $l'=2j-l$. The radial functions $F(r)$ and $G(r)$
satisfy the set of the radial Dirac equations:
\begin{eqnarray}
    c\frac{dG(r)}{dr}+\frac{c\kappa}{r}G(r) - (\varepsilon
+m_{\bar{p}}c^{2}-V(r))F(r) - \frac{\kappa_{0}}{2m_{\bar{p}}c}V^{'}(r)G(r)
= 0, \nonumber \\
    c\frac{dF(r)}{dr}-\frac{c\kappa}{r}F(r) + (\varepsilon
-m_{\bar{p}}c^{2}-V(r))G(r) + \frac{\kappa_{0}}{2m_{\bar{p}}c}V^{'}(r)F(r)
= 0.
    \label{radeq}
\end{eqnarray}

The phase shifts $\delta_{j,\,l}^{s}$ can be obtained from asymptotic
expansions of $F(r)$ and $G(r)$ as $r\rightarrow\infty$. An alternative
way to calculate $\delta_{j,\,l}^{s}$ is the variable phase
method \cite{Calogero,Babikov1,Grechukhin}. Within this approach, the
variable phase $\delta_{j,\,l}^{s}(r)$ is a solution of a first-order
differential equation:
\begin{eqnarray}
\frac{d}{dr}\delta_{l\pm 1/2,\,l}^{s}(p,r)=-\frac{(pr)^{2}}{c}
\nonumber  \\
\nonumber
\times\left(v(r)\sqrt{\frac{\varepsilon+m_{\bar{p}}c^{2}}
{\varepsilon-m_{\bar{p}}c^{2}}}\left[\cos\delta_{l\pm
1/2,\,l}^{s}(p,r)F^{\mathrm{C}}_{l}(pr) -\sin\delta_{l\pm
1/2,\,l}^{s}(p,r)G^{\mathrm{C}}_{l}(pr)\right]^{2} \right.\\
\nonumber +\left. v(r)\sqrt{\frac{\varepsilon-m_{\bar{p}}c^{2}}
{\varepsilon+m_{\bar{p}}c^{2}}}\left[\cos\delta_{l\pm
1/2,\,l}^{s}(p,r)F^{\mathrm{C}}_{l\pm1}(pr) -\sin\delta_{l\pm
1/2,\,l}^{s}(p,r)G^{\mathrm{C}}_{l\pm1}(pr)\right]^{2}\right. \\
\nonumber -\left. \frac{\kappa_{0}}{mc}V^{'}(r)\left[\cos\delta_{l\pm
1/2,\,l}^{s}(p,r)F^{\mathrm{C}}_{l}(pr) -\sin\delta_{l\pm
1/2,\,l}^{s}(p,r)G^{\mathrm{C}}_{l}(pr)\right]\right. \\
\times \left. \left[\cos\delta_{l\pm
1/2,\,l}^{s}(p,r)F^{\mathrm{C}}_{l\pm1}(pr) -\sin\delta_{l\pm
1/2,\,l}^{s}(p,r)G^{\mathrm{C}}_{l\pm1}(pr)\right]\vphantom{\sqrt{\frac{
\varepsilon+m_{\bar{p}}c^{2}}
{\varepsilon-m_{\bar{p}}c^{2}}}}\right)\,, \label{faza}
\end{eqnarray}
$F^{\mathrm{C}}_{l}(pr)$ and $G^{\mathrm{C}}_{l}(pr)$ being the regular
and irregular Dirac wave functions for the pure Coulomb potential,
respectively, and $v(r)$ is the short-range part of the scattering
potential,
\begin{equation}
 v(r) = V(r)+ \frac{Z}{r} = V_{\mathrm{n}}(r) + V_{\mathrm{U}}(r) +
V_{\mathrm{WK}}(r) + \frac{Z}{r}\,.
\end{equation}
Here $V_{\mathrm{n}}(r)$ is the electrostatic potential of interaction
with the finite nucleus and $V_{\mathrm{U}}(r)$ is the Uehling potential
which is given by the lowest-order term in the expansion of the
one-electron-loop vacuum polarization in powers of the Coulomb
electron-nucleus interaction. The Wichmann--Kroll potential
$V_{\mathrm{WK}} (r)$ accounts for the higher-order terms in the expansion
of the vacuum loop in powers of the Coulomb electron-nucleus interaction
\cite{Wichman56}. The explicit forms of the potentials $V_{\mathrm{n}}(r)$,
$V_{\mathrm{U}}(r)$ and $V_{\mathrm{WK}}(r)$ can be found elsewhere
\cite{Maiorova07,Maiorova08,Shabaev02,Mohr98,Artemmyev97}.

In order to calculate the differential cross section (\ref{dcs}), one has
to evaluate the quantities $A(\theta)$ and $B(\theta)$ in equations
(\ref{rjadA}) and (\ref{rjadB}) which contain infinite summations over
the angular momenta. For the slowly decaying (Coulomb-tail) interaction,
the series (\ref{rjadA}) is formally divergent because the phaseshifts
$\delta_{l\pm1/2,\,l}$ do not decrease with increasing $l$. For the pure
Coulomb potential, the regularization procedure was suggested by Mott
\cite{Mott32} who obtained also the first numerical results for the
relativistic Coulomb scattering. Detailed theoretical and numerical
studies of this problem can be found in a number of papers (see,
e.g., \cite{Dogget56,Sherman56,Gluckstern64}); it can be regarded
as well-understood and does not pose any challenge nowadays. Once the
scattering amplitudes $A^{\mathrm{C}}(\theta)$ and $B^{\mathrm{C}}(\theta)$
for the Coulomb field are computed, the amplitudes $A(\theta)$ and
$B(\theta)$ for the given potential $V(r)$ can be calculated as follows:
\begin{eqnarray}
A(\theta)= A^{\mathrm{C}}(\theta)-
\frac{1}{2\mathrm{i}p}\sum_{l=0}^\infty\left[(l+1)
\left(\exp(2\mathrm{i}\delta_{l+1/2,\,l}^{c})-\exp(2\mathrm{i}\delta_{l+1/2,\,l})
\right)\right. \nonumber \\
\left.+l\left(\exp(2\mathrm{i}\delta_{l-1/2,\,l}^{c})
-\exp(2\mathrm{i}\delta_{l-1/2,\,l})\right)\right]
P_{l}(\cos\theta),\label{rjadAp}
\end{eqnarray}
\begin{eqnarray}
B(\theta) = B^{\mathrm{C}}(\theta)-\frac{1}{2p}
\sum_{l=1}^\infty\left[\left(\exp(2\mathrm{i}\delta_{l+1/2,\,
l}^{c})-\exp(2\mathrm{i}\delta_{l+1/2,\, l})\right) \right. \nonumber \\
\left.
-\left(\exp(2\mathrm{i}\delta_{l-1/2,\,l}^{c})-\exp(2\mathrm{i}\delta_{l-1/2,\,l}
)\right) \right]P^{1}_{l}(\cos\theta).\label{rjadBp}
\end{eqnarray}

\section{Results and discussion}
We have computed the differential cross sections of relativistic antiproton
scattering by bare uranium nuclei in the energy range from 100~eV to
2.5~keV. The Coulomb glory effect can be estimated if we compare the
DCS in the backward direction with that of the non-relativistic Rutherford
scattering; the latter has a smooth minimum at $\theta=180^{\circ}$
irrespectively of the energy of the scattered particle. It is convenient
to define the scaled DCS which is being measured in the units of the
Rutherford cross section in the backward direction [$(Z/4E)^{2}$]:
\begin{equation}
 \frac{d\tilde{\sigma}}{d\Omega} =
 \left(\frac{4E}{Z}\right)^{2}\frac{d\sigma}{d\Omega} .
\label{scs}
\end{equation}
\begin{figure}

\begin{center}
\includegraphics[clip,width=0.7\columnwidth]{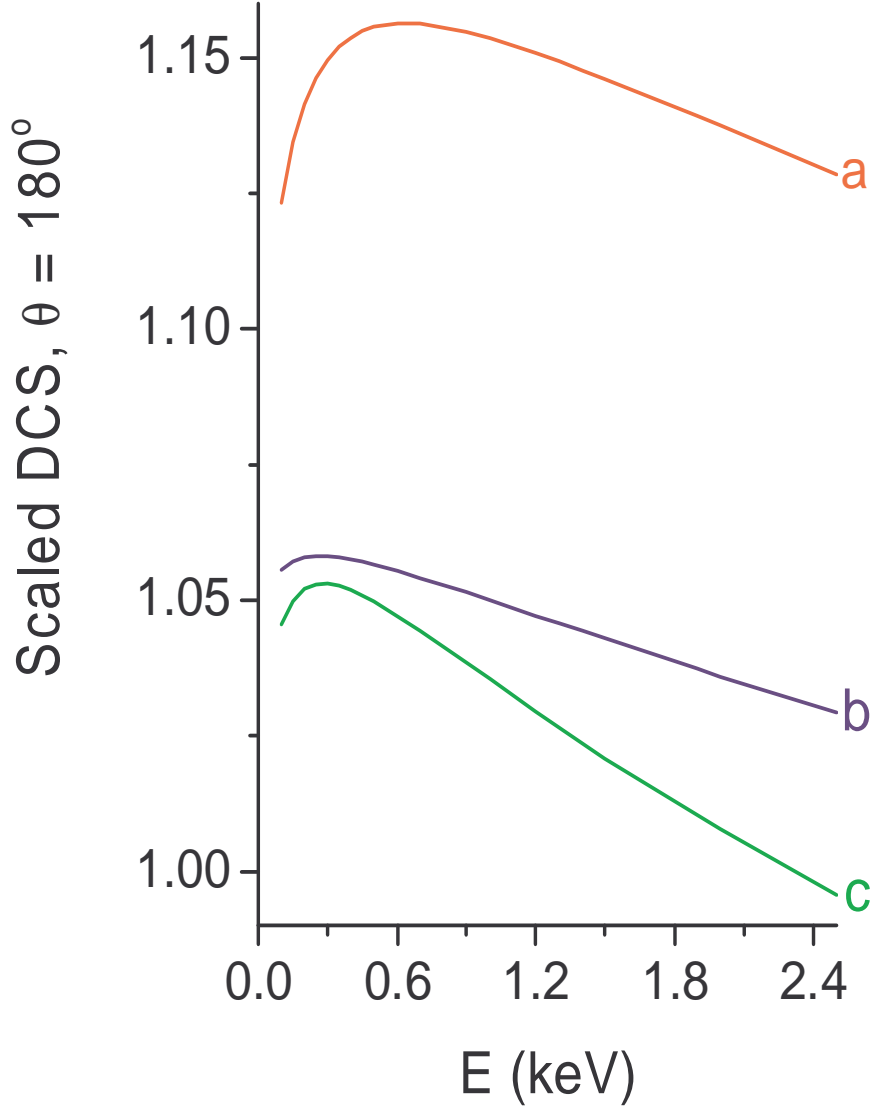}
\caption{\label{fig1} Scaled DCS $d\tilde{\sigma}/d\Omega$ at $\theta =
180^{\circ}$ as a function of the energy. (a) DCS for the total scattering
potential (relativistic); (b) DCS for the finite nucleus potential
only (relativistic); (c) DCS for the total scattering potential
(non-relativistic).}
\end{center}
\end{figure}
Using the scaled DCS, we can compare the results at different
energies and determine the energy domain with the largest Coulomb
glory effect. In figure \ref{fig1} we show the scaled DCS at
$\theta = 180^{\circ}$ as a function of the antiproton energy. The
maximum Coulomb glory effect is reached at the antiproton energy
of about 600~eV and amounts to 15.6\% (figure \ref{fig1}, curve
(a)). The main contribution comes from the VP potential, as the
DCS produced by the finite nucleus potential only (figure
\ref{fig1}, curve (b)) is about 10\% smaller and does not exhibit
a pronounced maximum. We note that the contribution from the
anomalous magnetic moment to $d\tilde{\sigma}/d\Omega$ is very
small and has the order of magnitude $\sim 10^{-4}$, that is in
agreement with a qualitative estimate given by Milstein
\cite{mil08}. With the help of the optical potential method
\cite{Batty97,Gal00}, we have also estimated the influence of the
strong interaction between the antiproton and the nucleus and
found it negligible. In the non-relativistic limit
$c\rightarrow\infty$, the present results coincide with the
previous calculations \cite{Maiorova08} which are represented by
curve (c) in figure \ref{fig1} \footnote{The curve (c) in figure 2
of \cite{Maiorova08} corresponding to the non-relativistic DCS for
the finite nucleus potential only appeared incorrect in the paper;
it must be shifted down by about 3-5\%. All other curves in that
figure are correct.}.
 As one can see, the relativistic
corrections are very important: they are responsible for overall
increase of the DCS by approximately 10\% and the shift of the
maximum to higher energies (from 300~eV to 600~eV). In figure
\ref{fig2}, we present the DCS dependence on the scattering angle
at the antiproton energy $E = 600$~eV corresponding to the
strongest Coulomb glory effect. The curve (a) shows the DCS
resulting from the full interaction while the curve (c) represents
the DCS produced by the finite nucleus potential only; the curve
(b) corresponds to the pure (point charge) Coulomb potential. The
relativistic contribution to the DCS at $\theta=180^{\circ}$ for
the pure Coulomb potential (figure \ref{fig2}, curve (b)) amounts
to 14.3\% at this energy, in good accordance with the value
obtained from the approximate equation (\ref{asimp}). An
oscillatory dependence of the DCS on the scattering angle has the
same nature as in the non-relativistic case where it appears due
to the interference between the Coulomb and short-range
contributions to the scattering amplitude \cite{semicl}; similar
oscillations for the relativistic Coulomb scattering were observed
in \cite{Fradkin64}. In the Coulomb glory case, the highest
maximum appears at $\theta=180^{\circ}$ (figure \ref{fig2}, curve
(a)), and its width is about $5^{\circ}$.

\begin{figure}
\includegraphics[clip,width=\columnwidth]{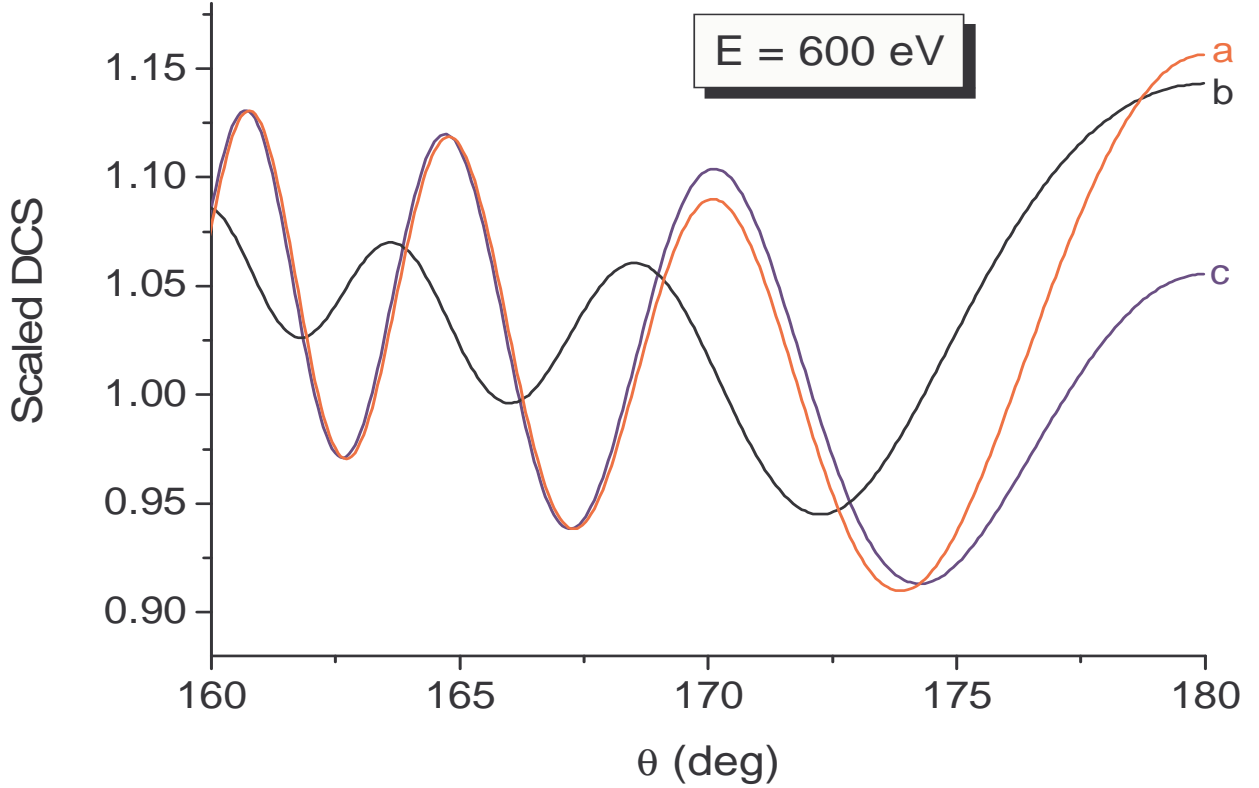}
\caption{\label{fig2} Scaled DCS $d\tilde{\sigma}/d\Omega$ (\ref{scs})
for the energy of the antiproton 600 eV. (a) DCS for the total scattering
potential; (b) DCS for the exact Coulomb potential (relativistic); (c)
DCS for the finite nucleus potential only.}
\end{figure}

In order to check that inelastic scattering channels do not mask
the Coulomb glory phenomenon, we have also evaluated the total
cross sections of such processes as radiative recombination (RR)
and annihilation of antiprotons. To compare the results, the cross
sections of the inelastic processes have been scaled in a similar
manner that was used for the elastic scattering DCS. Since the
width of the Coulomb glory maximum in the angular dependence of
the DCS is about $5^{\circ}$, we define the scale factor
$\sigma_{0}$ as the Rutherford DCS integrated over the same
angular domain in the vicinity of $\theta =180^{\circ}$:
\begin{equation}
\sigma_{0} =  \left(\frac{Z}{4E}\right)^{2}
\int\limits_{(180^{\circ}-\theta)\le5^{\circ}}\frac{d\Omega}{\sin^{4}
\theta/2 } .
\end{equation}
We have found that the total RR cross section (scaled by $\sigma_{0}$)
$\hat{\sigma}_{\mathrm{r}}$ does not exceed $ 10^{-4}$ and,
therefore, is small compared to the Coulomb glory effect. A rough estimate
of the scaled antiproton annihilation cross section
$\hat{\sigma}_{\mathrm{a}}$ can be obtained as follows:
\begin{equation}
 \hat{\sigma}_{\mathrm{a}} \sim \frac{\pi R_{\mathrm{n}}^{2}}{\sigma_{0}}
\frac{|\psi_{\mathrm{C}}(r=0)|^{2}}{|\psi_{\mathrm{f}}(r=0)|^{2}},
\end{equation}
where $R_{\mathrm{n}}$ is the nuclear charge radius,
$\psi_{\mathrm{C}}$ is the non-relativistic wavefunction of an
antiproton in the Coulomb field and $\psi_{\mathrm{f}}$ is the
wavefunction of a free antiproton. The results obtained by this
formula are presented in table \ref{tab1}. They show that the
antiproton annihilation should not mask the Coulomb glory effect.

\begin{table}
\caption{\label{tab1} Rough estimates of the scaled cross section of
the antiproton annihilation at different energies.}
%\begin{indented}
% \item[]
\begin{tabular}{@{}llllllll}
%\br
\hline
E (eV) & 100 & 600 & 1300 & 1900 & 2500 \\
%\mr
\hline
 $\hat{\sigma}_{\mathrm{a}}$ & 0.0006 & 0.01 & 0.03
 & 0.06 & 0.09 \\
 \hline
%\br
\end{tabular}
%\end{indented}
\end{table}

\section{Summary}
In this paper, the backward scattering of antiprotons by bare uranium
nuclei has been investigated in the framework of the relativistic theory in
the range of the antiproton kinetic energy 100~eV to 2.5~keV. The effects
due to vacuum polarization and finite size of the nucleus, as well as the
influence of the anomalous magnetic moment of the antiproton, have been
taken into account. It is the screening property of the one-loop VP
potential that is responsible for the Coulomb glory effect with the
prominent DCS maximum in the backward direction. Both the non-relativistic
and relativistic theories predict this maximum in some range of the
antiproton kinetic energy. The relativistic effects, however, significantly
alter the non-relativistic Coulomb glory picture. The kinetic energy
corresponding to the strongest effect is shifted to higher values (600~eV
versus 300~eV in the non-relativistic case), and the DCS in the vicinity
of $\theta =180^{\circ}$ becomes larger. We have also estimated the role of
 inelastic processes, such as the radiative
recombination and annihilation of antiprotons, and found that
they should not mask the Coulomb glory phenomenon.

\acknowledgments
We are thankful to Alexander Milstein for drawing
our attention to the importance of the relativistic effects and
for fruitful discussions. Valuable conversations with Yu.N. Demkov
are gratefully acknowledged. This work was supported by DFG (Grant
No. 436RUS113/950/0-1), by RFBR (Grant No. 07-02-00126a), and by
the Ministry of Education and Science of Russian Federation
(Program for Development of Scientific Potential of High School,
Grant No. 2.1.1/1136; Program ``Scientific and pedagogical
specialists for innovative Russia'', Grant No P1334). The work of
A.V.M. was also supported by the ``Dynasty'' foundation. V.M.S.
acknowledges financial support by the Alexander von Humboldt
Foundation.

\end{document}